\documentstyle[12pt]{article}
\newcommand{\be}{\begin{equation}}
\newcommand{\ee}{\end{equation}}
\newcommand{\bea}{\begin{eqnarray}}
\newcommand{\eea}{\end{eqnarray}}
\newcommand{\ba}{\begin{array}}
\newcommand{\ea}{\end{array}}

\font\mybb=msbm10 at 10pt
\def\bb#1{\hbox{\mybb#1}}

\def\bR {\bb{R}}

\begin{document}


\begin{titlepage}
\rightline{DAMTP-2006-80}
\rightline{ITFA-2006-32}
\rightline{hep-th/0609056}

\vfill

\begin{center}
\baselineskip=16pt
{\Large\bf Hamilton-Jacobi method for Curved Domain Walls\\}
\smallskip
{\Large\bf   and Cosmologies }
\vskip 0.3cm
{\large {\sl }}
\vskip 10.mm
{\bf ~Kostas Skenderis$^{*,1}$ and  Paul K. Townsend$^{\dagger,2}$}
\vskip 1cm
{\small
$^*$
Institute for Theoretical Physics, \\
University of Amsterdam,\\
Valckenierstraat 65, 1018 XE Amsterdam,\\
The Netherlands\\
}
\vspace{6pt}
{\small
$^\dagger$
Department of Applied Mathematics and Theoretical Physics,\\
Centre for Mathematical Sciences, University of Cambridge,\\
Wilberforce Road, Cambridge, CB3 0WA, U.K.
}
\end{center}
\vfill

\par
\begin{center}
{\bf
ABSTRACT}
\end{center}
\begin{quote}

We use Hamiltonian methods to study curved domain walls and cosmologies.
This leads naturally to first order equations for all 
domain walls and cosmologies foliated by slices of maximal symmetry.
 For Minkowski and AdS-sliced
domain walls (flat and closed FLRW cosmologies) 
 we recover a recent  result concerning their (pseudo)supersymmetry.
We show how domain-wall stability  
is consistent with the instability of adS vacua that violate the 
Breitenlohner-Freedman bound.  We also explore the relationship to  
Hamilton-Jacobi theory and compute the wave-function 
of a 3-dimensional closed universe evolving towards de Sitter spacetime.

\vfill
\vfill
\vfill

\vfill
 \hrule width 5.cm
\vskip
2.mm
{\small
\noindent $^1$ skenderi@science.uva.nl \\
\noindent $^2$ p.k.townsend@damtp.cam.ac.uk 
\\
}
\end{quote}
\end{titlepage}
\setcounter{equation}{0}
\section{Introduction}

Scalar fields are of particular interest in cosmology because a  time-dependent scalar  field is compatible with the isotropy and homogeneity  of spacelike hypersurfaces.  Scalar fields arise in the context of  domain  wall solutions of Einstein's equations for similar reasons; such solutions 
are particularly important for gravity/gauge theory dualities because they provide gravitational 
duals of quantum field theories. 
It is possible to consider both cases together  because to every domain wall solution of a model with a scalar-field potential-energy function $V$,  there is a corresponding cosmological solution of the same model but with potential-energy function $-V$, and vice-versa  \cite{Skenderis:2006jq}; in practice, this is achieved by  means of a sign $\eta$, such that $\eta=1$ for domain  walls and $\eta=-1$ for  cosmology.  Of course, cosmology is about evolution in time whereas domain walls are about evolution in space, but if the evolution of a scalar field $\sigma$ (in either time or space) is monotonic then one can instead consider the evolution, of the spacetime metric and any other scalar fields,
as a function of  $\sigma$. The time-dependence (if $\eta=-1$) or space-dependence  (if $\eta=1$) of 
both the spacetime metric and $\sigma$ (and any other scalar fields) is then found to be 
determined by a set of  {\it first-order} equations. 

This idea was first developed for the cosmological evolution of flat universes \cite{Salopek:1990jq}, where it was identified as an application of Hamilton-Jacobi (HJ) theory\footnote{A succinct explanation of the relevant computation can be found in Chapter 3 of the book by Liddle and Lyth \cite{LL}.}. 
A similar HJ approach for flat domain walls was initiated in \cite{deBoer:1999xf}. In the domain-wall case  there is a close connection to supersymmetry  because the HJ equation can be reduced to the
equation for the potential in terms of the superpotential, which is simply related to Hamilton's characteristic function.  The first-order equations mentioned above are the supersymmetry-preservation, or `BPS',  equations that arise as integrability conditions for the existence of a Killing spinor, and only very special solutions will be `supersymmetric' for a given superpotential.  However, as pointed out in \cite{Freedman:2003ax,Sonner:2005sj}  in the context of models with a single scalar field, any flat domain wall solution for which the scalar is strictly monotonic is `supersymmetric' in the above sense for 
{\it some} superpotential that solves the `reduced HJ equation'. Note that the `strictly monotonic' requirement excludes adS vacua, which  may always be viewed as flat domain walls \cite{Lu:1996rh} but which will be supersymmetric only if the Breitenlohner-Freedman bound \cite{Breitenlohner:1982bm} is satisfied. As we explain here, it also excludes domain wall solutions that are asymptotic to unstable adS vacua, thus resolving a potential paradox.

Of principal interest in \cite{Freedman:2003ax}, and some subsequent works 
\cite{Celi:2004st,Zagermann:2004ac}, were the conditions under which domain walls that are ``adS-sliced'', rather than flat, can be supersymmetric, in  the same sense of admitting Killing spinors for some choice of superpotential. It was known from earlier work \cite{Behrndt:2001mx,LopesCardoso:2001rt} 
that this requires a complex, or otherwise multi-component, superpotential. It was shown in \cite{Freedman:2003ax} that this superpotential must satisfy  a certain consistency condition and some complex superpotentials satisfying  this consistency condition were found, thus allowing a proof of classical stability for some special solutions (such as the Janus solutions \cite{Bak:2003jk} of the model with constant negative $V$). However,  these results left open the general problem, which was essentially solved in our previous paper \cite{Skenderis:2006jq}: almost any adS-sliced domain-wall 
solution determines a complex superpotential,  satisfying the required consistency condition, such that  the domain-wall solution is `supersymmetric', in the sense that it admits a Killing spinor. 
The `almost'  refers, firstly, to the same monotonicity requirement that  arises for flat domain walls. Secondly, there are difficulties that may arise in the multi-scalar case; we bypass these difficulties here by  restricting ourselves to single-scalar models.

One purpose of this paper is to show how the complex superpotential construction of our 
previous paper emerges naturally from the Hamiltonian formulation of the problem. As the superpotential is intrinsically complex for an adS-sliced wall, it is no longer so simply related to the characteristic function of HJ theory. Moreover,  HJ theory is applicable to domain wall solutions that  are dS-sliced (foliated by de Sitter spacetimes) as well as those that are adS-sliced, but only in the latter case does the construction of \cite{Skenderis:2006jq} define a superpotential.  This limitation is expected from  the connection to supersymmetry because the only dS-sliced `walls' that admit  Killing spinors  are dS-foliations of adS vacua. Thus, the construction of \cite{Skenderis:2006jq} cannot be viewed merely as an application of HJ theory. Because of the supersymmetry connection, one could consider it as a `square-root' HJ formalism.  However, it turns out that this formalism allows a construction that is slightly more general than that of \cite{Skenderis:2006jq}, and this generalized construction yields first-order equations for  dS-sliced walls too, although these  are not `BPS' equations (and they also differ from the first-order equations proposed in \cite{Afonso:2006gi}, which were based on the similar first-order equations for cosmologies found in \cite{Bazeia:2005tj}). 

As is well-known, HJ theory arises naturally in the semi-classical limit of quantum mechanics. 
As an application of our square root HJ  formalism we will obtain the wave-function for the closed $D=3$ `separatrix'  universe  found explicitly in \cite{Sonner:2005sj}, and which was shown there to correspond to the evolution 
from an Einstein Static universe to a de Sitter universe. 

Our starting point is a metric $g$ for a spacetime of dimension $D=d+1$, and a scalar field
 $\sigma$, with Lagrangian density
\be\label{Lstart}
{\cal L} = \sqrt{-\det g}\left[ R
  -\frac{1}{2}\left(\partial \sigma\right)^2
-V(\sigma)\right] \, ,
\ee
where $R$ is the curvature of $g$.
For the cosmological or domain wall solutions of interest here, the metric  can be put in the form
\be\label{ansatz} 
ds^2_D = \eta \left(f e^{\alpha\varphi}\right)^2 dz^2 + 
e^{2\beta\varphi}ds^2_d
\ee
where $ds^2_d$ is {\it either} (for $\eta=-1$) a space of constant curvature $k=-1,0,1$, {\it or} (for $\eta=1$) a spacetime of constant curvature $k=-1,0,1$,  and we have introduced the $D$-dependent constants 
\be
\alpha = (D-1) \beta\, , \ \qquad 
\beta =1/\sqrt{2(D-1)(D-2)}\, . 
\ee
We have allowed for an arbitrary function $f(z)$ in the ansatz (\ref{ansatz}) in order to maintain invariance under a change of the parameter $z$, which is a time parameter for $\eta=-1$ and a space parameter for $\eta=1$. The function $\varphi(z)$ determines the scale factor for the $d$-dimensional metric. 
The scalar field  $\sigma$ must be taken to be a function only of $z$ in order  to preserve the generic 
isometries of this spacetime metric. Note that $k=-1$ corresponds to adS-sliced domain walls  for $\eta=1$ and to open universes for $\eta=-1$, whereas $k=1$ corresponds to dS-sliced domain walls  for $\eta=1$ and to closed universes for $\eta=-1$.  
For this ansatz, the  field equations reduce to equations for the variables $(\varphi,\sigma)$ that are equivalent to the Euler-Lagrange  equations of the effective Lagrangian
\be\label{efflag}
L= \frac{1}{2} f^{-1} \left(\dot\varphi^2 - \dot\sigma^2\right) 
- f e^{2\alpha\varphi}\left(\eta V
- \frac{\eta k}{2 \beta^2} e^{-2\beta\varphi}\right) \, , 
\ee
where the overdot indicates differentiation with respect to $z$. 

\setcounter{equation}{0}
\section{Hamiltonian formulation}

Introducing the momentum variables $(\pi,p)$ conjugate to $(\varphi,\sigma)$, we can write the effective Lagrangian in the alternative Hamiltonian form
\be\label{HamLag}
L = \dot\varphi \pi + \dot\sigma p - f {\cal H}
\ee
where 
\be\label{hamorig}
{\cal H} = {1\over2}\left(\pi^2-p^2\right) + e^{2\alpha\varphi} \left(\eta V - {\eta k\over 2\beta^2} e^{-2\beta\varphi}\right)\, . 
\ee
and $f$ is a Lagrange multiplier (the `lapse' function) for the Hamiltonian constraint ${\cal H}=0$. 
The equations of motion are
\be\label{eqsofm1}
f^{-1}\dot\varphi = \pi\, ,\qquad f^{-1}\dot\sigma = -p
\ee
and
\be\label{momderiv}
f^{-1}\dot\pi = -2\alpha e^{2\alpha\varphi} \eta V + 
{k\eta\over 2\alpha\beta^2}e^{2(\alpha-\beta)\varphi}\, , \qquad
f^{-1}\dot p = - e^{2\alpha\varphi} \eta V'\, . 
\ee

We now observe that, for {\it arbitrary} complex function $Z(\sigma)$,
\bea\label{hamrewrite}
{\cal H} &\equiv& \frac{1}{2} \left\{\pi + 2\alpha e^{\alpha\varphi} \left[\frac{{\cal R}e \left(\bar Z Z'\right)}{|Z'|}\right]\right\}
 \left\{\pi - 2\alpha e^{\alpha\varphi} \left[\frac{{\cal R}e \left(\bar Z Z'\right)}{|Z'|}\right]\right\}
\nonumber\\
&-&\!\! \frac{1}{2}\left\{p+ 2e^{\alpha\varphi} |Z'|\right\}\left\{ p- 2e^{\alpha\varphi} |Z'|\right\}
- \frac{e^{2\alpha\varphi}}{2\beta^2}\left\{ \eta k \, e^{-2\beta\varphi} +
\left[ \frac{2\alpha\beta\, {\cal I}m \left(\bar Z Z'\right)}{|Z'|}\right]^2 \right\} \nonumber\\
&+& \!\! e^{2\alpha\varphi}\left\{\eta V - 2|Z'|^2 + 2\alpha^2 |Z|^2\right\}\, . 
\eea
All terms on the right hand side that involve $Z$ cancel, and one is then left with the original expression 
(\ref{hamorig}). Given any solution of the equations  of motion and the Hamiltonian 
constraint  for which $\sigma(z)$ is strictly monotonic, there exists an inverse function $z(\sigma)$ that allows any function of $z$ to be considered as a function of $\sigma$. We could then use the equations\footnote{A different relative sign leads to an inconsistency.}
\be\label{firstorder1}
\pi = \pm 2\alpha \ e^{\alpha\varphi}  \left[\frac{{\cal R}e \left(\bar Z Z'\right)}{|Z'|}\right]\, ,\qquad
p = \pm 2 \ e^{\alpha\varphi} |Z'|\, , 
\ee
to {\it define} the function $Z$. Actually,  as we will see, this defines
a 1-parameter family of functions $Z_u(\sigma)$, where $u$ is a real number.

If the equation for $p$ in (\ref{firstorder1}) is differentiated and the
equations of motion in (\ref{momderiv}) used, one finds a differential equation 
that can be once-integrated to yield
\be \label{pot}
\eta V = 2|Z'|^2 - 2\alpha^2 (|Z|^2 -u) 
\ee
where $u$ is an integration constant. Using this relation, and (\ref{firstorder1}), in (\ref{hamrewrite}), 
one arrives at the `reduced'  Hamiltonian constraint 
\be \label{ueqn}
\left[\frac{{\cal I}m \left(\bar Z Z'\right)}{|Z'|}\right]^2  +
\frac{\eta k}{4 \alpha^2 \beta^2} \, e^{-2\beta\varphi} =u\, .
\ee

If the equation for $\pi$ in (\ref{firstorder1}) is differentiated and the
equations of motion in (\ref{momderiv}) used, one finds that
\be \label{picomp}
{\cal I}m \left(\bar Z' Z''\right)  \,
{\cal I}m \left(\bar Z Z'\right) 
+ \frac{\eta k}{4 \alpha \beta}  e^{-2\beta\varphi} |Z'|^2 =0\, , 
\ee
and using  (\ref{ueqn}) to eliminate the $\varphi$ dependence we deduce that
\be
{\cal I}m \left(\bar Z' Z\right)
{\cal I}m \left(\bar Z' (Z'' + \alpha \beta Z) \right)  
= \alpha \beta u |Z'|^2\, . 
\ee
Recall that the equations of motion were assumed to be solved so that that 
these equations are {\it consequences} of the equations of motion and the 
Hamiltonian constraint; i.e.  {\it identities}. 

By combining the equation for $\pi$ in (\ref{firstorder1})  with (\ref{ueqn}), one can show that
\be\label{defining}
4\alpha^2  (|Z|^2 -u)=  \left(e^{-\alpha\varphi}\pi\right)^2 -
 {k\eta\over\beta^2}e^{-2\beta\varphi} \, . 
 \ee
We may use this equation, and the equation for $p$ in (\ref{firstorder1}), to obtain a concrete 
formula for $Z$. Writing 
\be
Z= W e^{i\theta}
\ee
one has
\be\label{halfway}
4W^2(\theta')^2 \equiv 4|Z'|^2 -4(W')^2 =  e^{-2\alpha\varphi}p^2 -4(W')^2\, , 
\ee
where the second equality relies on the expression for $p$ in (\ref{firstorder1}).  An expression for $W'$ can now be obtained by differentiating (\ref{defining}) and using the lemma
\be
\left[ \left(e^{-\alpha\varphi}\pi\right)^2 -
 {k\eta\over\beta^2}e^{-2\beta\varphi} \right]' = 2\alpha e^{-2\alpha\varphi} p \pi \, , 
 \ee
which is proved by using the equations of motion and the constraint. Using  the expression for $W'$ thus obtained in (\ref{halfway}), one finds an expression for $\theta'$. The net  result is that  $W$ and $\theta'$ are given by
\bea\label{om}
W_u(\sigma) &=& \frac{1}{2\alpha}\, 
\sqrt{(e^{-\alpha\varphi} \pi)^2  -
(k\eta/\beta^2)e^{-2 \beta \varphi}
+ 4 \alpha^2 u}\, , \\
\theta_u'(\sigma) &=& \pm \frac{\alpha (e^{-\alpha\varphi} p) 
\sqrt{4 \alpha^2 u 
-(k\eta/\beta^2) e^{-2 \beta \varphi}}}{  
(e^{-\alpha\varphi} \pi)^2 -(k\eta/\beta^2)
e^{-2 \beta \varphi} + 4 \alpha^2 u}\, . 
\label{thetaprime}
\eea
For $u=0$, and the gauge choice $f= e^{-\alpha\varphi}$, these equations reduce to those given in \cite{Skenderis:2006jq} when use is made of (\ref{eqsofm1});  as observed there, these equations  are the integrability conditions  for  {\it either} a  Killing spinor (if $\eta=1$) {\it or} a pseudo-Killing spinor 
(if $\eta=-1$). 

The ambiguity represented by the constant $u$ could have been anticipated from the fact that
the equations (\ref{firstorder1}) do not fix $Z$ uniquely because there
may be different complex functions with the same norm of $Z'$ and
real part of $\bar Z' Z$. Indeed let us define
\be 
\zeta = \omega e^{i \psi}
\ee
and set
\be \label{zeta}
\omega = \sqrt{|Z|^2 -u} \, , \qquad 
\psi' = \frac{\sqrt{\left({\cal I}m \bar Z Z'\right)^2 - u |Z'|^2}}{|Z|^2 -u}\, . 
\ee
One may check that the following relations hold
\bea
&& |\zeta'|^2=|Z'|^2, \qquad {\cal R}e\, 
\bar{\zeta} \zeta = {\cal R}e \bar{Z} Z \nonumber \\
&& |\zeta|^2 = |Z|^2 - u, \qquad 
\left({\cal I}m\, \bar \zeta \zeta'\right)^2 =  \left({\cal I}m\, \bar Z Z'\right)^2 - u |Z'|^2\, . 
\eea
So the form of the first order equations (\ref{firstorder1}) remains unchanged
but  we now have 
\be\label{preHJ}
\eta V - 2 |\zeta'|^2 + 2 \alpha |\zeta|^2 =0 \, , \qquad 
{\cal I}m \left(\bar \zeta \zeta' \right)
{\cal I}m \left[ \bar \zeta' \left(\zeta'{}' +\alpha\beta \zeta\right)\right]=0\, , 
\ee
and\footnote{The sign of the exponent on the right hand side 
corrects a typographical error in \cite{Skenderis:2006jq}.}
\be
\left[ \frac{{\cal I}m 
\left(\bar \zeta \zeta'\right)}{|\zeta'|}\right]^2  = -
\frac{\eta k}{4 \alpha^2 \beta^2} \, e^{-2\beta\varphi}\, . 
\label{ketaeq} 
\ee
In other words, (\ref{zeta}) maps the $u\neq 0$ cases to the $u=0$ case, as long as $k\eta\le0$.
When $k=0$,  (\ref{ketaeq}) implies that
\be\label{keqzero}
{\cal I}m \left(\bar \zeta \zeta' \right) =0\, ,  \qquad \left(k=0\right), 
\ee
and the complex superpotential is real\footnote{For positive $u$ there is an alternative
description in terms of the complex superpotential  $Z = \sqrt{u}(1 + i \tan \theta)$, but 
nothing is gained by this.}. When $k \eta=-1$  the superpotential 
is intrinsically complex and satisfies
\be\label{concon}
{\cal I}m \left[ \bar \zeta' \left(\zeta'{}' 
+\alpha\beta \zeta \right)\right]=0\, , \qquad (\eta k=-1). 
\ee

We have now recovered the results of  \cite{Skenderis:2006jq}, which apply only when $k\eta\le0$. 
Indeed, when $k \eta>0$,  equation (\ref{ueqn}) implies that 
\be
\left({\cal I}m \bar Z Z'\right)^2 - u |Z'|^2 <0\, , 
\ee
which invalidates the map  (\ref{zeta}). Thus, the constant $u$ yields nothing new as long as 
$k\eta\le0$, but is essential to the validity of the formalism when $k\eta=1$. We will  conclude
with a simple example that illustrates the formalism for all values of $k\eta$.

\subsection{An illustrative example}

Consider a $D=5$ theory with potential 
\be
V(\sigma)= 3 \eta \cosh^2 \left(\frac{c \sigma}{\sqrt{6(c^2-\eta k)}} \right)\, 
\left[c^2 + 3 \eta k - 4 c^2 \tanh^2 \left(\frac{c \sigma}{\sqrt{6(c^2-\eta k)}}\right)\right]\, , 
\ee
for constant $c$. Choosing the gauge  $f=e^{-\alpha \varphi}$, 
one can show that 
the equations of motion are solved \cite{Gremm:2000dj}, for 
any value of $\eta k$, 
by\footnote{This solution has a naked 
singularity at $z=\pm \pi/(2 c)$.}
\bea \label{solu}
\varphi(z) &=& 2 \sqrt{6} \log c z \\
\sigma(z) &=& \frac{1}{c} \sqrt{6(c^2-\eta k)} \, 
\log \left(\frac{1+\tan c z/2}{1-\tan c z/2}\right)\, . \nonumber
\eea
The complex superpotential $Z$ can be derived following the 
steps we outlined:
\bea
W_u(\sigma) &=& 3 \sqrt{c^2  \sinh^2 \hat{\sigma}
- \eta k \cosh^2 \hat{\sigma}
+\frac{1}{9} u} \\
\theta'_u(\sigma) &=& \mp
\frac{\sqrt{(c^2 - \eta k)} \, 
\cosh \hat{\sigma}
\sqrt{\frac{1}{9} u
- k \eta \cosh^2 \hat{\sigma}}
}{\sqrt{6} \left(c^2  \sinh^2 \hat{\sigma}
- \eta k \cosh^2 \hat{\sigma}
+\frac{1}{9} u\right)} \nonumber 
\eea
where 
\be
\hat{\sigma}= c \sigma/\sqrt{6(c^2-\eta k)}\, . 
\ee
One may check that  the potential is indeed given by (\ref{pot}) and the solution (\ref{solu})
solves the first order equations (\ref{firstorder1}), for any value of $\eta k$.

\setcounter{equation}{0}
\section{Domain-wall stability}

As shown in \cite{Skenderis:2006jq}, and earlier works cited there, the first-order equations found  for $k\eta\le0$ are `BPS' equations that arise as integrability conditions for the existence of either a  Killing  spinor (if $\eta=1$) or a pseudo-Killing spinor (if $\eta=-1$).  In the ($\eta=1$) domain-wall case 
it follows that `almost all'  flat or adS-sliced domain walls are `supersymmetric'.  The importance of supersymmetry in this (`fake') sense lies in its  connection to classical stability. This has long been appreciated for adS-vacua, where the criterion for classical stability is that the scalar field mass satisfies the Breitenlohner-Freedman (BF) bound \cite{Breitenlohner:1982bm,Mezincescu:1984ev}, because this bound can be shown to follow from the assumption that the adS vacuum is `supersymmetric' \cite{Boucher:1984yx,Townsend:1984iu}. We will now review this argument, partly to fix notation 
but also to explain how we now see this old result in a new light. Moreover, we present it for 
either choice of the sign $\eta$, although the physical interpretation is straightforward only for 
$\eta=1$. 

We begin by observing that the relevant equations, 
in the gauge $f = e^{-\alpha\varphi}$,
are the second-order equation of motion
\be
\ddot\sigma =-\alpha\dot\varphi\dot\sigma + \eta V'\, ,
\ee
and the constraint
\be\label{twoeq}
\dot\varphi^2 -\dot\sigma^2+ 2\eta V = {k\eta\over \beta^2} e^{-2\beta\varphi}\, . 
\ee
Differentiation of this constraint  yields a combination of the $\sigma$ equation of motion and the  second-order equation for $\varphi$.  Maximally-symmetric vacua are solutions of these equations for which $\sigma$ is a constant, which we may take to be zero and which must be an extremum of $V$. The equation of $\varphi$ is then
\be\label{solvephi}
\dot\varphi^2 + 2\eta V_0 = {k\eta\over \beta^2} e^{-2\beta\varphi}\, ,  
\ee
where $V_0=V(0)$. For all other solutions, $\dot\sigma$ is not identically zero.  We will  suppose that $V$ has an extremum at $\sigma=0$ with $\eta V_0 <0$, so for  $\eta=1$ the vacuum is $adS_D$ and for $\eta=-1$ the vacuum is  $dS_D$.  In this case $V$ has the Taylor expansion
\be\label{taylor}
 V= -{\eta\over 2\beta^2 \ell^2} + {1\over2}  m^2 \sigma^2 + {\cal O}\left(\sigma^3\right)
\ee
where $\ell$ is the (a)dS radius and $m$ the mass of the scalar field fluctuation.  We are now in a position to state the (generalized) BF bound:
\be\label{genBF}
\eta\,  m^2 \ge   - {(D-1)^2\over 4\ell^2}\, .
\ee
{}For $\eta=1$ this is the standard BF bound, which states that $m$ cannot be ``too tachyonic''.  For $\eta=-1$, it is a `cosmological bound'  that sets an upper limit to $m$. Of course, in neither case  is the bound absolute; it simply serves to separate (a)dS vacua with different physical features.  In the $\eta=-1$ case, the relevant feature has a simple physical interpretation in terms of a damped harmonic oscillator: the bound requires the curvature of the potential to be small enough that the motion is overdamped; if the bound is violated then the motion is underdamped and the approach to equilibrium involves an infinite number of oscillations. These oscillations force $\dot\sigma$ to change sign an infinite number of times. 

To derive the bound, we begin by assuming that  the potential $V$ can be written in terms of a  real function $W$ such that 
\be \label{superpotV}
\eta V = 2\left[(W')^2 - \alpha^2 W^2\right] \, . 
\ee
We must  suppose that $V$ has an extremum in order for there to exist an (a)dS vacuum. Note that
\be
\eta V' = 4W'\left[ W'{}' -\alpha^2 W\right]
\ee
so that extrema of $V$ correspond either to extrema of $W$ or to solutions of $W'{}'=\alpha^2 W$. For 
$\eta=1$ we shall say that the adS vacua arising from $W'=0$ are `supersymmetric'. Similarly, for $\eta=-1$ we shall say that the dS vacua arising from $W'=0$ are pseudo-supersymmetric.  For such a vacuum we have (for either choice of  $\eta$)
\be
\alpha^2 \frac{V''}{V} = -2\frac{W'{}'}{W^2}
\left[W'{}'  - \alpha^2 W\right] \, .  
\ee
This is a quadratic equation for $W'{}'$, with real roots iff 
\be\label{altbound}
\frac{V'{}'}{V} \le \frac{\alpha^2}{2} = {(D-1)\over 4(D-2)}\, . 
\ee
{}For $V$ of the form (\ref{taylor}), this yields the bound (\ref{genBF}).  Although this derivation of  the BF bound was given (for $\eta=1$) in  \cite{Boucher:1984yx,Townsend:1984iu}, the main point there was to derive the expression (\ref{superpotV}) from the requirement of positive energy by means of a Nester-Witten type of  argument; it was assumed from the outset that the superpotential was extremized at the adS vacuum. Our current viewpoint is rather different: the expression  (\ref{superpotV}) is now viewed not as a constraint on $V$ but rather as a differential equation to be solved for $W$ given $V$, and the non-trivial assumption is that  $W'=0$.  The superpotential $W$ exists regardless of whether the BF bound  is satisfied, but if it is not satisfied then $W' \neq 0$ and  the (a)dS vacuum is not  (pseudo)supersymmetric.   For an adS vacuum that {\it is} supersymmetric, by this definition 
of `supersymmetry', the arguments of \cite{Boucher:1984yx,Townsend:1984iu} 
imply that it is stable. 

Notice that the supersymmetry condition $W'=0$ for adS vacua is just the 
special case of one of the  first-order `BPS' equations for domain-walls, 
so one may imagine that the Nester-Witten-type proof
used in \cite{Boucher:1984yx,Townsend:1984iu} to prove the positivity of the 
energy in adS vacua with $W'=0$ will extend to supersymmetric domain-wall 
spacetimes. Indeed it was noticed in \cite{Skenderis:1999mm}
that flat domain walls satisfying the first order equations
also saturate the positive energy bound derived in \cite{Townsend:1984iu}, 
which implies that any such wall has the lowest possible energy among all
solutions with the same asymptotics (although there may be other domain walls 
with the same energy). Such a positive energy theorem was discussed in detail for flat and
adS-sliced walls that  are asymptotically adS in \cite{Freedman:2003ax}.
A scrutiny of the arguments presented there reveals (although we will not attempt 
a detailed proof)  that they apply  to any domain wall spacetime, irrespective of 
asymptotic conditions. Given this, any flat or adS-sliced wall that is supersymmetric must also be classically stable.

Thus, not only are `almost all' flat or adS-sliced domain-wall spacetimes `supersymmetric', they are also (as a consequence) classically stable. This conclusion leads to  a paradox when one considers that potentials $V(\sigma)$ that admit an unstable adS vacuum will also admit flat and adS-sliced domain-wall solutions that are asymptotic to this unstable vacuum. How can a stable domain-wall solution be asymptotic to an unstable adS vacuum? As we will now show, the resolution of this paradox lies in the  qualification `almost' as applied to the above conclusion concerning the `supersymmetry' of domain walls. Recall that this qualification is required because we assumed strict monotonicity of the scalar field $\sigma$. As observed in \cite{Skenderis:2006jq}, zeros of $\dot\sigma$ correspond to branch points of a multi-valued superpotential, and a different branch will generally be required on either side of a zero of $\dot\sigma$. Nevertheless, any domain-wall solution for which the zeros of $\dot\sigma$ are isolated can be said to be ``piecewise-supersymmetric'', and one would expect this to be sufficient for stability. If, however, there is an accumulation of the zeros of $\dot\sigma$ then the argument for supersymmetry, and hence for stability, fails. As we will now show, any unstable adS vacuum is an accumulation point for zeros of $\dot\sigma$ on any domain-wall spacetime that is asymptotic to it.

We need only consider the equations for domain walls in the neighborhood of the adS vacuum. As $\sigma\equiv0$ in this vacuum we may assume that both $\sigma$ and $\dot\sigma$ are small and neglect terms of quadratic or higher order in these variables.  Using the form of the potential (\ref{taylor})
the equation for $\sigma$ becomes
\be\label{sigmaeq}
\ddot\sigma + \alpha\dot\varphi\dot\sigma - \eta m^2\sigma = 0\, , 
\ee
and the constraint becomes
\be
\dot\varphi^2 = \frac{1}{(\beta\ell)^{2}} 
+ {k\eta\over \beta^2} e^{-2\beta\varphi}\, , 
\ee
which is just (\ref{solvephi}).  This equation is easily solved and the solution may be used in (\ref{sigmaeq}) to get a linear differential equation for $\sigma$, which we need to analyze for 
$k=0$ and $k=-1$. As we are now discussing domain walls we should set $\eta=1$ but, in view of possible future applications to cosmology, we will continue to allow for either sign in the equations to follow. 

Let us first consider $k=0$; in this case $\dot\varphi= (\beta\ell)^{-1}$ and hence 
\be
\ddot\sigma + (D-1)\ell^{-1} \dot\sigma -\eta m^2\sigma =0\, .
\ee
This has a solution of the form $\sigma \propto e^{\gamma z}$ for real $\gamma$ iff 
the generalized BF bound  (\ref{genBF}) is satisfied. This solution approaches
the adS vacuum solution (for $\eta=1$) with $\sigma\equiv0$ monotonically as either $z\to\infty$ or as 
$z\to -\infty$. If the bound is {\it not} satisfied then the solution has the form
\bea
\sigma &=& e^{-(D-1)z/(2 \ell)}
\left[ A \cos(\omega z) + B\sin(\omega z)\right]\, , \nonumber\\
\omega &\equiv&  \frac{1}{2 \ell} \sqrt{-4\eta m^2\ell^2 -(D-1)^2}
\eea
This still approaches the adS vacuum, as $z\to\infty$, but $\sigma(z)$ is no longer monotonic. 
 In fact $\dot\sigma=0$ an infinite number of times, with the adS solution being an accumulation point for zeros of $\dot\sigma$.  
As just mentioned, the proof that a flat domain wall  is supersymmetric assumes that $\dot\sigma$ has no zeros. It it has isolated zeros then one can still prove supersymmetry piecewise, but not even this is possible if the zeros are not isolated. A domain wall solution that is asymptotic to an an unstable adS vacuum has non-isolated zeros because this solution comes arbitrarily close to the accumulation point of zeros of $\dot\sigma$ at the adS vacuum. If this region is excluded then we recover piecewise supersymmetry. 

For the $k=-1$ case we have
\be
\dot\varphi = (\beta\ell)^{-1} \tanh \left(z/\ell\right)
\ee
and hence
\be
\ddot\sigma + (D-1)\ell^{-1}  \tanh\left(z/\ell\right)\, \dot\sigma -\eta m^2 \sigma =0\, .
\ee
Note that $\tanh(z/\ell)<1$, so the $\dot\sigma$ coefficient is bounded for all $z$. Also, as $|z|\to\infty$, we recover the $k=0$ case, up to the sign of $z$, so the conclusions of the $k=0$ case still apply. 

We have now resolved the paradox by demonstrating that  there is no contradiction between the possibility of an unstable adS vacuum and the supersymmetry of the $k\le0$ domain walls asymptotic to it; the
supersymmetry of such walls is only local, and piecewise, and cannot be
extended to the full domain-wall solution.
However, it is of interest to consider  the $k=1$ case too. In this case, 
\be
\dot\varphi =(\beta\ell)^{-1} \coth \left(z/\ell\right)
\ee
and hence 
\be
\ddot\sigma + (D-1)\ell^{-1} \coth\left(z/\ell\right)\, \dot\sigma -\eta m^2\sigma =0\, .
\ee
Note that $\coth(z/\ell)$ blows up as $z\to 0$. The singularity  is not relevant if $\sigma$ and $\dot\sigma$ are not small as $z\to\infty$ because the solution would then have moved away from the neighborhood of the adS vacuum but it could happen that $\dot\sigma \to 0$ as $z\to 0$. In this case the relevant equation near $z=0$ would be
\be \label{z=0}
\ddot\sigma + {(D-1)\over 2z} \dot\sigma -\eta\, m^2 \sigma =0\, . 
\ee
This equation has a regular singularity at $z=0$ with indices $0, -(D-3)/2$. The latter possibility yields a solution singular at $z=0$ but the former one yields a solution for which $\dot\sigma\to 0$ as $z\to 0$, as assumed, but $\sigma$ tends to a non-zero constant for this solution, which is therefore not in the neighborhood of the adS vacuum solution. Thus, even for $k=1$ the solution can approach the adS vacuum only as $|z|\to\infty$ and then the conclusions of the $k=0$ case again apply.

\setcounter{equation}{0}
\section{Hamilton-Jacobi Theory}

For any mechanical model with $2(n+1)$ phase-space variables $(Q,P)$ and reparametrization invariant action, the phase-space Lagrangian takes the form
\be\label{PQLag}
L= \dot Q  P - f \, {\cal H}(Q,P)\, . 
\ee
The Hamilton-Jacobi equation for this model is the partial differential equation
\be
{\cal H}(Q,\partial S/\partial Q) =0
\ee
for Hamilton's ``characteristic''  function $S(Q)$. Given a solution of this equation, one 
defines 
\be
P= \frac{\partial S}{\partial Q}\, . 
\ee
A particular solution $S(Q,{\bf p})$ of the HJ equation involves $n$ independent integration constants\footnote{It is $n$ rather than $n+1$ because the
constraint allows one of $n+1$ integration constants, typically the 
energy,  to be expressed in terms of the rest.} ${\bf p}$, and each such solution yields an $n$-dimensional family of solutions of the Euler-Lagrange equations of (\ref{PQLag}).  These solutions,  in reparametrizaton invariant form,  are 
found by solving the equations
\be\label{qeq}
\frac{\partial S}{\partial {\bf p}} = {\bf q}
\ee
for $n$ constants ${\bf q}$. As an illustration, consider the relativistic particle of mass $m$ with
${\cal H}=p^2+m^2$.  The  HJ equation is solved by
\be
S= {\bf p} \cdot {\bf x} - E t\, , \qquad E\equiv \sqrt{|{\bf p}|^2 +m^2}\
\ee
for integration constants ${\bf p}$, and (\ref{qeq}) then yields
\be
{\bf x} = {\bf q} + \left[{\bf p}/E\right] t\, . 
\ee
This reproduces the usual solution in the gauge $\dot t=1$. 

This HJ formalism can be applied to the effective Lagrangian (\ref{HamLag}) for domain walls and cosmologies. The HJ equation is
\be  \label{HJeqn}
\left(\frac{\partial S}{\partial\varphi}\right)^2 - \left(\frac{\partial S}{\partial\sigma}\right)^2
+2e^{2 \alpha\varphi}\left[ \eta V - \frac{\eta k }{2\beta^2} e^{-2\beta\varphi}\right] =0
\ee
and the momenta conjugate to $(\varphi,\sigma)$ are given by
\be
\pi = \frac{\partial S}{\partial\varphi}\, ,\qquad p= \frac{\partial S}{\partial\sigma}\, . 
\ee
We could now proceed as in the above example. However, we instead follow 
an alternative `square-root'  procedure that is based on  factorization of the Hamiltonian 
constraint, as in (\ref{hamrewrite}). Specifically, we may rewrite the HJ 
equation, for arbitrary complex function $Z(\sigma)$, as
\bea\label{hamrewrite2}
{\cal H} &\equiv& \frac{1}{2} \left\{\frac{\partial S}{\partial\varphi} 
+ 2\alpha e^{\alpha\varphi} \left[\frac{{\cal R}e \left(\bar Z Z'\right)}{|Z'|}\right]\right\}
 \left\{\frac{\partial S}{\partial\varphi} - 2\alpha e^{\alpha\varphi} \left[\frac{{\cal R}e \left(\bar Z Z'\right)}{|Z'|}\right]\right\}
\nonumber\\
&-&\!\! \frac{1}{2}\left\{\frac{\partial S}{\partial\sigma} + 2e^{\alpha\varphi} |Z'|\right\}\left\{ \frac{\partial S}{\partial\sigma} -  2e^{\alpha\varphi} |Z'|\right\}
- \frac{e^{2\alpha\varphi}}{2\beta^2}\left\{ \eta k \, e^{-2\beta\varphi} +
\left[ \frac{2\alpha\beta\, {\cal I}m \left(\bar Z Z'\right)}{|Z'|}\right]^2 \right\} \nonumber\\
&+& \!\! e^{2\alpha\varphi}\left\{\eta V - 2|Z'|^2 + 2\alpha^2 |Z|^2\right\}\, . 
\eea

We begin our analysis with the observation that
\bea
\dot S &=& \dot\sigma p + \dot\varphi \pi \nonumber\\
&=& \pm 2e^{\alpha\varphi} \left\{ \dot\sigma |Z'| +
\alpha\dot\varphi \left[\frac{{\cal R}e \left(\bar Z Z'\right)}{|Z'|}\right]\right\} \nonumber\\
&=& \frac{d}{dz}\left\{ \pm 2e^{\alpha\varphi} \left[\frac{{\cal R}e \left(\bar Z Z'\right)}{|Z'|}\right]\right\}  \mp \eta k \, \frac{\dot\sigma e^{(\alpha-2\beta)\varphi}}{2\alpha\beta |Z'|}
\eea
where we have used (\ref{firstorder1}) to get to the second line, 
and integrated by parts and used (\ref{picomp})
to get to the last line. We may now use (\ref{eqsofm1}) and
(\ref{firstorder1}) to deduce
\be
\dot S =  \frac{d}{dz}\left\{ \pm 2e^{\alpha\varphi} \left[\frac{{\cal R}e \left(\bar Z Z'\right)}{|Z'|}\right]\right\} 
+  \frac{\eta k}{\alpha\beta} f e^{2(\alpha-\beta)\varphi}\, . 
\ee
Thus
\be\label{hamchar1}
S= \pm 2e^{\alpha\varphi} \left[\frac{{\cal R}e \left(\bar Z Z'\right)}{|Z'|}\right]
\ + \ \frac{k\eta}{\alpha\beta} \int^{z(\sigma)} e^{2(\alpha-\beta)\varphi(\xi)}f(\xi)\,  d\xi
\ee
Note the appearance of the integrated term for $k\ne0$. 
The value of this term depends on the choice of the function $f$. 
We may always choose
\be\label{nice}
f = e^{-2(\alpha-\beta)\varphi}\, , 
\ee
in which case
\be\label{hamchar2}
S(\sigma,\varphi) = \pm 2e^{\alpha\varphi} \left[\frac{{\cal R}e \left(\bar Z Z'\right)}{|Z'|}\right]
+ \frac{k \eta}{\alpha\beta} z(\sigma)\, . 
\ee
This is the characteristic function for our problem. Notice that it takes the same form whether expressed in terms of $Z$ or in terms of $\zeta$, these functions being related by (\ref{zeta}).

One may verify that 
\be\label{momcons}
\frac{\partial S}{\partial\varphi} = \pm  2\alpha \ e^{\alpha\varphi}  
\left[\frac{{\cal R}e \left(\bar Z Z'\right)}{|Z'|}\right]\, , \qquad
\frac{\partial S}{\partial\sigma} = \pm 2 \ e^{\alpha\varphi} |Z'| \, . 
\ee
The first of these relations follows by inspection. To prove the second relation we first use
the lemma 
\be
\left( \frac{{\cal R}e \left(\bar Z Z'\right)}{|Z'|} \right)' = 
|Z'| - \frac{1}{|Z'|^3} {\cal I}m (Z'' \bar{Z}') {\cal I}m (Z' \bar{Z})
=|Z'| + \frac{\eta k}{4 \alpha \beta} \frac{e^{-2\beta\varphi} }{|Z'|} 
\ee
where the second equality uses (\ref{picomp}). Then we use 
\be
z' (\sigma) =\frac{1}{\dot\sigma} = 
\mp \frac{ e^{-\alpha\varphi}}{2 f |Z'|}  = \mp \frac{ 
e^{(\alpha- 2\beta)\varphi}}{2 |Z'|}\, , 
\ee
where the last equality uses (\ref{nice}).  Using (\ref{momcons}) in (\ref{hamrewrite2}) we have
the `reduced' HJ equation
\be\label{HJcan}
\eta V - 2|Z'|^2 + 2\alpha^2 |Z|^2 = 
\frac{1}{2\beta^2}\left\{ \eta k \, e^{-2\beta\varphi} +
\left[ \frac{2\alpha\beta\, {\cal I}m \left(\bar Z Z'\right)}{|Z'|}\right]^2 \right\}\, . 
\ee

Let us first recall that for $k=0$ we have $Z=W$, a real function 
of $\sigma$, so the right hand side  of (\ref{HJcan}) vanishes and the 
`reduced' HJ equation becomes
\be\label{superpoteq}
\eta V= 2\left[(W')^2 - \alpha^2 W^2\right]\, , 
\ee
which is the equation for the potential in terms of the superpotential $V$. 
Given a solution of this equation for $W$ we then find $S$ from
\be
S= \pm 2 e^{\alpha\varphi} W\, . 
\ee
Recall that $S$ (which is also the on-shell value of the action)
and the canonical momenta $p$ and $\pi$
have a direct interpretation in the AdS/CFT correspondence. 
The latter are the 1-point functions of the corresponding
dual operators \cite{deBoer:1999xf,Papadimitriou:2004ap}.
The fact that $S$ turns out to be a local function of  $\sigma$ means that 
one can add a finite local boundary counterterm such that the renormalized
on-shell action vanishes \cite{Bianchi:2001de}. It then follows that the 
ground state energy of the dual QFT is zero, in agreement with supersymmetry,
and the 1-point function of the operator dual to $\sigma$ is also zero.

{}For $k\eta=-1$ we have a rather more complicated state of affairs.  
Recall that $S(\varphi,\sigma)$ was found by using the field equations; 
its arguments are boundary values of functions $\varphi(z)$ and $\sigma(z)$ that solve these equations. As we saw earlier, the field equations imply that the right hand side of  (\ref{HJcan}) is zero. 
So when $k\eta=-1$ the `reduced' HJ equation is
\be\label{genHJ1}
\eta V= 2\left[|Z'|^2 - \alpha^2 |Z|^2\right]\, , 
\ee
but, as we saw earlier, this must be solved subject to the constraint 
\be\label{genHJ2}
{\cal I}m \left[ \bar Z' \left(Z'{}' +\alpha\beta Z\right)\right]=0\, . 
\ee
Thus, we really have two ``HJ equations'' to solve for the complex function 
$Z$, which then yields the characteristic function $S$ via the formula 
(\ref{hamchar1}), or (\ref{hamchar2}). Regarding holography now,  
we note that one may still add a local counterterm to cancel the first term in the right hand side of 
(\ref{hamchar2}) but the second term is a non-local 
function of $\sigma$ so it cannot be removed. So the 
ground state energy is still zero but the 1-point function of the operator
dual to $\sigma$ is in general non-zero\footnote{In the context of the Janus
solution mentioned earlier, this behavior is expected since the dual theory is 
(conjectured to be) a defect CFT.}. 

Finally, for $k \eta=1$ one must have $u \neq 0$. In this case the `reduced' HJ equations 
are
\bea
&&\eta V = 2|Z'|^2 - 2\alpha^2 (|Z|^2 -u)\, ,  \\
&&{\cal I}m \left(\bar Z' Z\right)
{\cal I}m \left(\bar Z' (Z'' + \alpha \beta Z) \right)  
= \alpha \beta u |Z'|^2\, . \nonumber
\eea
The above comments on holography in the $\eta k =-1$ case apply also to this case.

\subsection{A semi-classical wave-function}

The connection of HJ theory to semi-classical quantum mechanics is standard. The wave function 
$\Psi$ obeys the equation
\be\label{WdW}
{\cal H}\left(\varphi,\sigma, \hat\pi,\hat p\right) \Psi =0\, , 
\ee
where
\be
\hat\pi = -i \frac{\partial}{\partial\varphi}\, ,\qquad 
\hat p = -i \frac{\partial}{\partial\sigma}\, .  
\ee
In the semi-classical limit one has $\Psi = \exp(iS)$ where $S$ is Hamilton's characteristic function, satisfying the HJ equation. As we have seen, 
one can take $S$ to be of the form 
(\ref{hamchar1}) and we must then solve {\it either} for $W$ from (\ref{superpoteq}), if $k=0$, {\it or} for $Z$ from (\ref{genHJ1}) and (\ref{genHJ2}). As our formalism is new only for $k\eta \neq 0$,
we shall illustrate it by a determination of a wave-function for a particular exact  $k\eta=-1$ solution of the $D=3$ model with $V=-\eta$ \cite{Sonner:2005sj}. The solution was found in the gauge $f =e^{-\varphi}$, which coincides with (\ref{nice}) for $D=3$ since $\alpha=2\beta=1$, and is
\be
e^\varphi = 1+ e^{\sqrt{2}\, z}\, ,\qquad e^{-\sigma} = 1+ e^{-\sqrt{2}\, z}
\ee
where $1-e^\sigma$ is positive.  For $\eta=-1$ this represents the evolution from a $D=3$ Einstein Static Universe in the far past to $dS_3$ in the far future. For $\eta=1$, it is the ``separatrix wall'' solution that interpolates between a linear dilaton  $adS_2\times \bR$ solution and $adS_3$. Given this solution, one can use the construction of \cite{Skenderis:2006jq} that we have explained in detail here to compute the complex superpotential $Z=W\exp(i\theta)$. The result was given in \cite{Skenderis:2006jq}:
\bea
W &=& \sqrt{1-e^\sigma + \frac{1}{2}e^{2\sigma}} \nonumber\\
\theta &=& {\rm arctan}\,  \left[ e^{-\sigma} \sqrt{2(1-e^\sigma)}\right]  + \frac{1}{\sqrt{2}} \log \left(\frac{1-\sqrt{1-e^\sigma}}{1+ \sqrt{1-e^\sigma}}\right)\, . 
\eea

For the case under study, the characteristic function $S$ is given by 
\be\label{hamchar3}
S(\sigma,\varphi) = 2 e^\varphi 
\left[\frac{WW'}{\sqrt {(W')^2 + W^2(\theta')^2}} \right]
-2 \, z(\sigma)\, ,  
\ee
and we compute that
\be
S(\sigma,\varphi)  = \sqrt{2} e^{\varphi + \sigma} 
+ \sqrt{2}\, \log\left( e^{-\sigma}-1\right)\, . 
\ee
As  $z\to\infty$ we have $e^\sigma\approx 1$ and $\varphi\approx \sqrt{2}\, z$, and hence
\be
S \sim \sqrt{2} e^{\varphi} \to \infty \qquad (z\to \infty).
\ee
As $\varphi$ is rapidly increasing the wave-function is rapidly oscillating and the semi-classical approximation should be a good one. In contrast, as $z\to -\infty$ we have $e^\varphi\approx1$ and $\sigma \approx -\sqrt{2}\, |z|$, so 
\be
S\sim \sqrt{2}  \, (-\sigma) \to \infty \qquad (z\to -\infty)
\ee
but the phase is only weakly oscillating and we should therefore expect quantum corrections to be important. In the cosmology context, we have found the semi-classical wavefunction for a $D=3$ universe that evolves from a static quantum mechanical universe to a semi-classical inflating de Sitter universe. 

\section{Conclusions}

In this paper we have used Hamiltonian methods to study domain-walls and cosmological solutions of gravity coupled to a scalar field $\sigma$ with potential $V(\sigma)$, with the focus on solutions that admit a foliation with slices of maximal symmetry. For such solutions there is a domain-wall/cosmology correspondence  \cite{Skenderis:2006jq}:  for every domain-wall solution  there is a corresponding cosmological solution  with  $V\to -V$. For either domain walls or cosmologies, the scalar field can be chosen as the independent variable, provided it is strictly monotonic, which is the condition that $\dot\sigma\ne0$, where the overdot indicates a derivative with respect to the original independent (space or time) variable. Subject to this condition, we have  shown that any domain-wall or cosmological solution solves a set of first-order equations  associated with a complex function $Z(\sigma)$. 

In the case of domain walls, first-order equations arise naturally from the requirement of supersymmetry or, more generally, `fake' supersymmetry, and in these cases the function $Z$ can be interpreted as a superpotential. With the exception of de Sitter-sliced domain walls, the first-order equations found here coincide with those found in \cite{Skenderis:2006jq} and in the earlier work on this topic cited there. In the case of flat or closed cosmologies, the first-order equations coincide with those derived in \cite{Skenderis:2006jq} from the requirement of pseudo-supersymmetry. Here we have shown that there are similar first-order equations governing dS-sliced domain walls (and open cosmologies) that are {\it not} (pseudo)supersymmetric, although one cannot expect these equations to have a similar `BPS' interpretation as integrability conditions for the existence of (pseudo)Killing spinors. 

The importance of `fake' supersymmetry is that it is essentially equivalent to stability. Thus, 
all  flat or anti-de Sitter-sliced domain walls for which $\sigma$ is strictly monotonic are stable.
Anti-de Sitter vacua can be viewed as special cases of domain walls but, as $\dot\sigma\equiv0$ 
in this case, they need not be stable; stability in this case depends on whether the Breitenlohner-Freedman  bound on the scalar field mass is satisfied. The potential instability of anti-de Sitter vacua 
 is puzzling in the domain wall context because any anti de Sitter vacuum will be an asymptotic limit of some flat (and anti-de-Sitter sliced) domain walls. We have resolved this puzzle here by showing that an unstable anti-de Sitter vacuum is an accumulation point for zeros of $\dot\sigma$ for any domain wall (flat or otherwise) that is asymptotic to it.  An  interesting question raised by these results is the fate of unstable anti-de Sitter vacua, and the domain-walls that are asymptotic to them. A natural guess is that these solutions decay via a time-dependent  process to some  `near-by'  stable domain wall, and it would be interesting to see whether there exist  time-dependent solutions that realize such a decay process. 

The cosmological side of this story raises a related set of questions. We have seen  that the Brietenlohner-Freedman bound  translates in the cosmology setting to  an upper bound on the mass of the scalar. 
Below the bound the scalar field is `overdamped'  and approaches its equilibrium value monotonically. Above the bound it is `underdamped' and oscillates about its equilibrium value as this value is approached. This behavior has been noticed in the cosmological  literature but to our knowledge its physical significance (if any) has not been discussed. One would expect oscillations to lead to particle production, and hence possibly to an instability. 

 In the case of flat domain-walls and cosmologies, for which the superpotential $Z$ is real,  it has long been known that the first order equations arise naturally from  an application of Hamilton-Jacobi theory. In this case $Z$ is proportional to Hamilton's characteristic function $S$, and the Hamilton-Jacobi equation for $S$ can be reduced to the equation relating the potential to the superpotential. This is rather remarkable, even mysterious,  when one considers the generality of the Hamilton-Jacobi formalism. Here we have extended the Hamilton-Jacobi analysis  to curved domain-walls and cosmologies. In this case $Z$ is intrinsically complex and its relation to Hamilton's  characteristic function, which retains its interpretation as the on-shell value of the action,  is necessarily  more complicated. Moreover, the single `reduced' Hamilton-Jacobi equation is now replaced by a pair of
differential equations for  the complex function $Z$.

As an application to quantum mechanics of our Hamilton-Jacobi formalism, we computed the semi-classical wave-function for  a  (1+2)-dimensional  closed universe for which the classical solution 
(which describes the evolution from a static universe to an inflating de Sitter universe) is known exactly.  For this example, the semi-classical approximation is valid at late times but not at early times, for which
a full quantum treatment  is needed.  It would be interesting to extract the physics associated 
with this wavefunction, and to generalize the computation to higher dimensions.

\bigskip
\noindent
{\bf Acknowledgments.} PKT thanks the EPSRC for financial support. 
KS is supported by NWO via the Vernieuwingsimplus grant 
``Quantum gravity and particle physics''.

\end{document}